\documentclass[12pt]{article}
\usepackage{amsthm,amsfonts}
\usepackage{amsmath}
\usepackage{graphicx,color}
\usepackage[a4paper,top=2.54cm,bottom=2.54cm,left=2.54cm,right=2.54cm]{geometry}
\usepackage{enumerate}

\begin{document}
\newtheorem{remark}{Remark}
\newtheorem{experiment}{Experiment}

\begin{center}
\large{\textbf{Bayesian Approach to Handling Informative Sampling}}
\end{center}

\begin{center}
{\textbf{Research Project}}\\
\hfill \break
{\textbf{Anna Sikov}}\\
{\textbf{Institute of Mathematics and Statistics}}\\
{\textbf{University of Sao Paulo}}
\end{center}
\hfill \break
\hfill \break

\section{Resume} 
In the case of informative sampling the sampling scheme explicitly or implicitly depends on the response variable. As a result, the sample distribution of response variable cannot be used for making inference about the population. In this research I investigate the problem of informative sampling from the Bayesian perspective. Application of the Bayesian approach permits solving the problems, which arise due to complexity of the models, being used for handling informative sampling. The main objective of the research is to combine the elements of the classical sampling theory and Bayesian analysis, for identifying and estimating the population model, and the model describing the sampling mechanism. Utilizing the fact that inclusion probabilities are generally known, the population sum of squares of the models residuals can be estimated, implementing the techniques of the sampling theory. In this research I show, how these estimates can be incorporated in the Bayesian modeling and how the Full Bayesian Significance Test (FBST), which is based on the Bayesian measure of evidence for precise null hypothesis, can be utilized as a model identification tool. The results obtained by implementation of the proposed approach to estimation and identification of the sample selection model seem promising. At this point I am working on methods of estimation and identification of the population model. An interesting extension of my approach is incorporation of known population characteristics into the estimation process. Some other directions for continuation of my research are highlighted in the sections which describe the proposed methodology.   

\section{Introduction} 
Inference from a sample to the whole population usually employes estimation of the model, holding in the population based only on the sample measurements. In many practical situations sample selection probabilities depend directly on the values of the variable which is being investigated (outcome variable) even after conditioning on the covariates (auxiliary variables). This typically occurs when the sample selection process uses design variables, which are correlated with the outcome variable. In this case the sampling design is informative and the distribution of the measurements corresponding to the units belonging to the selected sample is different from the distribution of the measurements corresponding to the units in the population. If the sampling design is informative, the use of standard methods of estimation that ignore sample selection process may result in large biases. Hence, one of the most important problems, facing the analyst is selection an appropriate model underlying the sample selection process, and testing  ignorability of the sampling selection process. The approaches to handling informative sampling, which are proposed in the literature generally focus on estimation of unknown parameters, leaving the two defined problems almost unaddressed. The methods proposed in the literature, typically use either classical inference procedures, such as maximum likelihood estimation, based on the approximation of the model holding for the sample measurements (Pfeffermann et al., (1998), Pfeffermann and Sverchkov (2003)), or weighting procedures, which use either reciprocals of sampling probabilities (Binder (1983), Skinner et al. (1989), Pfeffermann (1993, 1996)), or their modifications, which are designed to reduce the variances (Pfeffermann and Sverchkov(1999), Beaumont (2008),  Kim and Skinner (2013)). The sampling probabilities are generally assumed to be known for the sampled units. 
Application of the maximum likelihood approach requires modeling of the population distribution and specification of the parametric form of the conditional expectations of the sample selection probabilities, given the outcome variable and the covariates. Two these models define the model, holding in the sample. In this case hypothesis testing can be carried out by application of the log-likelihood test statistics, the score test or the Wald test, however, these tests generally unstable, or computationally complicated. In the second case one can utilize the tests, proposed by Pfeffermann and Sverchkov (1999) or by Pfeffermann (1993). 
\par
In this article we apply a Bayesian approach to estimation of unknown parameters and hypothesis testing. In principle, the likelihood, based on the sample distribution, can be imbedded within a Bayesian model, and inference can be drawn using the MCMC techniques. The problem with this approach is that prior information on unknown parameters is generally unavailable. On the other hand, if improper priors are imposed, the resulting posterior distributions may likewise be improper. In addition, the resulting models are generally very complicated, which results in an extremely slow convergence of the MCMC process even for the models, containing a small number of unknown parameters. Another restriction of utilizing the sample likelihood is that in some cases the model may be unidentifiable (see Section 3.1).
%(see Pfeffermann et al. (1998)). 
As with the case of the methods, based on the sample likelihood, we assume in this article a parametric model on the population distribution, which models the process generating the population measurements, and a parametric form of the expectations of the sample selection probabilities, given the outcome variable and covariates. Assuming that both processes are independent, and that the sampling probabilities are observed for all the sampled units, we propose to estimate the population model and the expectations of the sample selection probabilities, based on the Hovitz-Tompson estimators for the population sums of squares of the residuals of the models. The derivatives of Hovitz-Tompson estimators with respect to unknown parameters provide statistics for the Bayesian inference. Conditional distribution of these statistics, given unknown parameters, can be approximated as multivariate normals, since they constitute sums of random variables. Imposing a flat multivariate normal prior on the vector of unknown parameters permits application of the MCMC techniques and therefore, estimation of unknown parameters. The main advantage of application of the Bayesian approach is that it permits solving the problems of hypothesis testing and model identification. For hypothesis testing we utilize the Full Bayesian Significance Test (FBST), introduced by Pereira and Stern (1999). The FBST is based on the Bayesian measure of evidence, favoring the null hypothesis. This measure is computed using the draws from the posterior distribution. The evidence value estimates the probability of tangential set of points having posterior density values higher than the supremum of the posterior density under the null hypothesis. The proposed test rejects the null hypothesis if this probability is high. The other important problem is identification of a model, describing the sample selection process, since statistical inference in this case typically requires modeling the expectations of the sampling probabilities given the covariates and the outcomes, the forms of which are seldomly known (Pfeffermann et al (1998), Pfeffermann and Sverchkov (1999)). 

\section{Summary of the Proposed Approaches} 
\subsection{Estimation Methods} 
In this section we provide a brief description of the problem of informative sampling and present the main findings in this area. Let $Y_i$ denote the value of the outcome variable $Y$, associated with unit $i$, belonging to a sample $S$, drawn from a finite population $U$, and let $X$ and $V$  denote the vectors of auxiliary variables, associated with unit $i$, such that $dim(X_i)=p$ and $dim(V_i)=q$. We assume that the population values are independent realizations from a distribution with probability density function $f_p(y_i | x_i, \theta)$, where $\theta$ is an unknown parameter, and that the sample selection mechanism employed a design variable $Z$, so that the probability of the units to be selected is proportional to the value of this variable (or to some function of this variable). Throughout this paper we assume that the model holding in the population is the normal linear regression model,  

\begin{center}
\begin{equation}
y_i=x_i \beta+ \epsilon_i, \epsilon_i \sim N(0,\sigma^2).
\end{equation}
\end{center}
		
where $\theta=(\beta,\sigma^2)$.

Let $\pi_i$ denote the sample inclusion probability of unit $i$, and $I_i$ denote the sampling indicator, which takes the value $1$ if unit $i$ was selected to the sample and $0$ othervise. Therefore, the distribution, holding in the sample for the unit $i$ is a conditional distribution, $f_s(y_i | x_i, v_i)= f(y_i | x_i, v_i, I_i=1)$, which, following Pfeffermann et al., (1998), can be presented as

		\begin{center}
		\begin{equation}
		f_s(Y_i | x_i, v_i)=f_p(Y_i | x_i) \frac{P(I_i=1 | y_i, v_i)}{P(I_i=1 | x_i, v_i)}
		\end{equation}
		\end{center}
where $P(I_i=1 | x_i, v_i )= \int f_p(Y_i | x_i) P(I_i=1 | y_i, v_i) dy_i$
Noting that $P(I_i=1 | y_i, v_i)= \int P(I_i=1 | x_i, v_i, \pi_i)f(\pi_i | y_i, v_i) d\pi_i= E_p(\pi_i | y_i, v_i)$, the relationship (2) can be also written in the form:
		\begin{center}
		\begin{equation}
		f_s(Y_i | x_i, v_i)=f_p(Y_i | x_i) \frac {E_p(\pi_i | y_i, v_i)  }  {  E_p(\pi_i | x_i, v_i)}
    \end{equation}
		\end{center}
where the index $p$ of the expectation indicates that it is computed with respect to the population distribution $f_p(y_i | x_i)$. If the inclusion probabilities are proportional to the values of the design variable $Z$, the equation (2) takes the form 
		%\begin{center}
		%\begin{equation}
		$$f_s(Y_i | x_i, v_i)=f_p(Y_i | x_i) \frac {E_p(Z_i | y_i, v_i)  }  {  E_p(Z_i | x_i, v_i)}$$
    %\end{equation}
		%\end{center}
Note that the model, holding in the sample, $f_s(y_i | x_i, v_i)$ is fully determined by the population model, $f_p(y_i | x_i)$ and by $E_p(\pi_i | y_i, x_i)$. Note also that if selection probabilities do not depend on the value of the outcome variable $y_i$, then the distributions in the population and in the sample coincide. Utilizing the result obtained by Pfeffermann et al., (1998), which states that under common sampling designs with unequal sampling probabilities, when the population measurements are independently drawn from some distribution, the sample measurements are asymptotically independent as the population size increases, a sample likelihood can be specifies as 
		\begin{center}
		\begin{equation}
		L_{Samp}=\prod_{i=1}^n f_p(Y_i | x_i;\theta)  \frac {E_p(\pi_i | y_i, v_i; \gamma)  } { E_p(\pi_i | x_i, v_i; \theta, \gamma)},
    \end{equation}
		\end{center}
where the unknown parameters $\theta$ and $\gamma$ are indexing the sample model and a model underlying a sample selection mechanism correspondingly. The authors note that the functional form of the expectations, $E_p(\pi_i | y_i, x_i)$ is not necessarily known, and therefore, must be approximated. They propose two possible approximations:
		\begin{center}
		\begin{equation}
		E_p(\pi_i | y_i, v_i) \approx \sum_{j=0}^J A_j y_i^j+ h( v_i)    
		\end{equation}
		\end{center}

where $h(v_i )= \sum_{p=1}^{m} \sum_{m=1}^{K(p)}B_{kp} x_{ip }^k$ and $\left\{ A_j\right\}$ and $\left\{ B_{kp}\right\}$ are unknown parameters, and 
		\begin{center}
		\begin{equation}
		E_p(\pi_i | y_i, v_i) \approx \exp [\sum_{j=0}^J A_j y_i^j+ h( v_i)]
		\end{equation}
		\end{center}
The authors point out that under approximation (6) the resulting sample distribution may be not identifiable. For example, if the population distribution is normal, $Y_i | x_i \sim N(\beta_0+x_i^t \beta, \sigma^2)$ and $E_p(\pi_i | y_i, x_i)= \exp(A_1 y_i +g(x_i))$ for some function $g(x)$ then, by (3) it follows that the distribution of $Y_i | x_i$, holding in the sample is $N((\beta_0+A_1 \sigma^2)+x_i^t \beta, \sigma^2)$, which is not identifiable. 
In order to avoid identification problems the authors propose to split the estimation process into two steps, where in the first step the parameters $\left\{ A_j\right\}$ and $\left\{ B_{kp}\right\}$ are estimated from the observed probabilities $\pi_i$, and in the second step the parameters, indexing the population model are estimated from the likelihood, defined by (4) ,with the parameters $\left\{ A_j\right\}$ and $\left\{ B_{kp}\right\}$ substituted by their estimates. Apart from solving identifiability problem, when using the approximaion (6), this approach is designed to ease the computation process, which can be cumbersome for both approximations, (5) and (6), if the number of parameters, indexing the resulting sample model is large. In order to test sampling ignorability one can use the log-likelihood test statistic, obtained by application of  the generalized likelihood ratio test, however this test is applicable if all the parameters, $\theta$ and $\gamma$ are estimated from the likelihood (4). If parameters $\gamma$ are estimated from the observed probabilities $\pi_i$, one can use a Score or a Wald test. The described two step procedure was applied by Pfeffermann and Sverchkov (1999) for estimation of the parameters of linear regression model with normal error terms. Noting that, 
$E_s ( w_i | y_i,v_i )= \frac {1} {E_p(\pi_i | y_i, v_i ) }$, where $w_i=\frac {1}{\pi_i}$, the authors propose at the first step to identify and to estimate the unknown parameters $\left\{ A_j\right\}$ and $\left\{ B_{kp}\right\}$ by regressing $w_i$ against $(x_i, y_i)$. 
If the assumptions of normality is dropped then the parameters $\beta$ can alternatively be estimated as a solution of
		\begin{center}
		\begin{equation}
		\sum_{i=1 }^n {\frac {w_i}{{\hat{E}}_s(w_i | x_i, v_i)}} (y_i-x_i \beta) x_{si} = 0, \quad s=1,...,p
		\end{equation}
		\end{center}
where ${\hat{E}}_s(w_i | x_i, v_i)$ denote the estimates of the expectations,$E_s(w_i | x_i, v_i)$, obtained by regressing $w_i$ against $(x_i, y_i)$ (see Pfeffermann and Sverchkov, 1999 for details). For extension of the approach to the generalized linear models see Pfeffermann and Sverchkov (2003). 
The approach based on minimization of (7) can be viewed as a modification of probability weighting methods, which use census estimating equations for deriving unknown parameters $\theta$:
		\begin{center}
		\begin{equation}
		\sum_{i=1 }^N  h( y_i,x_i; \theta )=0		
		\end{equation}
		\end{center}
Since the measurements outside of the sample are unobserved, the left handside of (8) is replaced by its Horvitz-Thompson estimator 
		\begin{center}
		\begin{equation}
		\sum_{i=1 }^n w_i h( y_i,x_i; \theta )=0		
		\end{equation}
		\end{center}
If $h( y_i,x_i; \theta )= \frac {\partial log f_p(Y_i | x_i; \theta) }{\partial \theta}$ then the estimation method is known in the literature as pseudo likelihood estimation method, which was introduced by Skinner et al (1989). Under the model (1) the estimators $\widetilde\beta$ are the solutions of the equations
		\begin{center}
		\begin{equation}
		\sum_{i=1 }^n w_i(y_i-x_i \widetilde{\beta} ) x_{si} = 0, \quad s=1,...,p
		\end{equation}
		\end{center}
In order to test ignorability of the sampling mechanism, one can use the test proposed by Pfeffermann and Sverchkov (1999). Their test employes the relationship,
		\begin{center}
		\begin{equation}
		E_p( Y_i | v_i)= \frac{E_s(w_i Y_i | v_i) }{E_s(w_i|v_i)}
		\end{equation}
		\end{center}
(see Skinner, 1994). Denoting by $\epsilon_i= y_i-E_p(Y_i | x_i )$ the regression residuals associated with the unit $i$, one can test the hypothesis of the form $E_p( \epsilon_i^k )=E_s( \epsilon^k_i ), \quad k=1,2... ,$  which by (11) is equivalent to testing the hypothesis $Corr_s( \epsilon_i^k, w_i )=0, \quad k=1,2...$, where $Corr_s$ denotes the correlation under the sample distribution. The authors point out that it generally suffices to test the first 2-3 correlations. This approach can be applied to testing ignorability of the sampling mechanism, however, it does not permit identification of the model, underlying the sample selection mechanism. There exist a few other approaches to testing ignorability of the smpling mechanism. These approaches are generally based on the difference between the estimators of the regression coefficients under the assumed model and the model under ignorable sampling design. See Pfeffermann (1993) for discussion. 

\subsection{The FBST}
As said previously, implementation of the Bayesian approach permits application of the FBST for solving the problem of hypothesis testing and model identification. This can be carried out by determining whether the fitted model contains non-significant parameters. In our application this reduces to testing nested models, where the more complex model is tested versus the model under the null hypothesis, obtained by setting the coefficients of some group of variables to zero. Therefore, the FBST can be used as an identification tool for selecting the model in the family of the nested models which best fits the data. 
\par
Let us consider a standard parametric statistical model, i.e., for an integer $m$, $\theta \in \Theta \subseteq \Re^n$ is the parameter, $g(\theta)$ a prior probability density over $\Theta$, $x$ is the observation (a scalar or a vector), and $L_x(\theta)$ is the likelihood generated by the data $x$. After the data $x$ have been observed, the sole relevant entity for the evaluation of the Bayesian evidence value, $ev$, is the posterior probability (density) for $\theta$ given $x$, denoted by 

		\begin{center}
		\begin{equation}
		g_x (\theta)= g(\theta | x ) \propto g( \theta) L_x (\theta)		
		\end{equation}
		\end{center}

We are restricted to the case where the posterior probability distribution over $\Theta$ is absolutely continuous, that is $g_x(\theta)$ is a density over $\Theta$. We are focusing on testing of sharp hypothesis, which state that $\theta$ belongs to a sub-manifold $\Theta_H$ of smaller dimension than $\Theta$. For simplicity we use $H$ for $\Theta_H$ in sequel. Let $r(\theta)$ be a reference density on $\Theta$ such that the function $s(\theta )=g_x(\theta ) / r(\theta )$ is called the “relative surprise”. Now consider a sharp hypothesis $H: \theta \in \Theta_H$. Let $\widetilde{s}= \sup\limits_{H}s(\theta)$ and $T= \left \{ \theta \in \Theta: s(\theta)>\widetilde{s} \right\} $. The Bayesian evidence value agains $H$ tis defined as the posterior probability of the tangential set, i.e., 

		\begin{center}
		\begin{equation}
		\overline{ev} = P(\theta \in | x) = \int_{T} g_x( \theta ) d\theta  
		\end{equation}
		\end{center}
Note that the evidence value, supporting $H$,  $1-\overline{ev}$, is not an evidence against the alternative hypothesis $A$. Equivalently, $ev$ does not constitute an evidence in favor of $A$, although it is against $H$. The FBST rejects $H$ whenever $ev$ is small.      

\section{Proposed Approach}
\subsection{Estimation Equations and Bayesian Model}
As noted by Pfeffermann et al., (1998), the situation, often occurring in practice is where the only design information, available to the analyst is the vector of sample inclusion probabilities of the sample units, and possible, also the sample values of $Z$. In this research we assume that we observe both, sample inclusion probabilities and the values of the design variables, and that sampling probabilities are proportional to the corresponding values of the design variable. 
Let the population outcomes follow the model defined by (1) and assume for simplicity that

		\begin{center}
		\begin{equation}
		E(\pi_i | v_i, y_i) =v_i \gamma_v+\gamma_y y_i, \quad i=1,...,n		
		\end{equation}
		\end{center}
We also assume that the population realizations of $Z$ are independent between the units.  
%It must be emphasized that since the sample selection probabilities $\pi_i$ are proportional to the corresponding value of $Z_i$,  
%$E(\pi_i | v_i, y_i) =v_i {\widetilde{\gamma}}_v+{\widetilde{\gamma}}_y y_i$  for $i=1,...,n$.
\begin{remark}[]
The defined sample selection model is a special case of a model (5) where $J=1$ and $h()$ is a linear function of the covariates. However our approach can be easily extended to the situations where $J>1$ and $h()$ is a polynomial of an order $m$, for some $m>1$. The method presented below can also be utilized if conditional expectation of the sample selection probabilities is given by (6).
\end{remark}
In this research we differentiate between randomization based inference over all possible sample selections, and inference, based on both, randomization and model distributions. The former distribution is underlying classical survey sampling inference, while the later distribution takes also into account the process of generating the values of the finite population. The randomization distribution accounts only for the process of sample selection, holding the population values fixed, therefore inference based on this distribution is restricted to the populations, which are similar to the population of the study (see Pfeffermann, 2011 for discussion). On the other hand,  inference, based on both, randomization and model distributions requires strong assumptions on the design variables. For example, in our case assuming a parametric form only for the conditional expectation of $\pi_i$ given the values  $y_i$ and $v_i$ will not be sufficient. In this research we focus on a randomization based inference. 
\par Let $W(\gamma)$ define the population sum of squares of the regression residuals of the model for the design variable, 

		\begin{center}
		\begin{equation}
		W(\gamma) = \sum_{i=1}^N (\pi_i-(\gamma_v v_i+\gamma_y y_i))^2
		\end{equation}
		\end{center}
where $\gamma= (\gamma_v, \gamma_y)$ and $\gamma_v= (\gamma_1,..., \gamma_q)$.
If all population measurements were available, the value of $\gamma$ could be obtained as a solution of the estimation equations,

		\begin{center}
		\begin{equation}
		\begin{split}
		\frac {\partial W(\gamma)} {\partial \gamma_{v_s}}  =  \sum_{i=1 }^N (\pi_i-( \gamma_v v_i+\gamma_y y_i)) v_{si} =0,\quad s=1,...,q;  \\
		\frac {\partial W( \gamma)} {\partial \gamma_{y}}  =  \sum_{i=1 }^N (\pi_i-(\gamma_v v_i+\gamma_y y_i)) y_i =0
		\end{split}
		\end{equation}
		\end{center}
In real situation statistical inference can only be based on the available sample measurements. Denote by $\widetilde{W}(\gamma)$ the Horvitz-Thompson estimator of $W(\gamma)$, based on the observed sample $S$. Then
		\begin{center}
		\begin{equation}
		\widetilde{W}(\gamma) = \sum_{i \in S} \frac {(\pi_i-(\gamma_v v_i+\gamma_y y_i))^2} {\pi_i}
		\end{equation}
		\end{center}

Let $$\frac {\partial \widetilde {W}(\gamma)} {\partial \gamma_{v_l}}  =  \sum_{i \in S} \frac {(\pi_i-( \gamma_v v_i+\gamma_y y_i)) v_{li}} 
		{\pi_i}, \quad l=1,...,q;$$ and  $$\frac {\partial \widetilde {W}(\gamma)} {\partial \gamma_{y}}  =  \sum_{i \in S} { \frac{(\pi_i-(\gamma_v v_i+\gamma_y y_i)) y_i}{\pi_i}}.$$

Note that the equations specified above, can be alternatively expressed as 
$$\frac {\partial \widetilde {W}(\gamma)} {\partial \gamma_{v_l}}  =  \sum_{i=1}^N \frac {(\pi_i-(\gamma_v v_i+\gamma_y y_i)) v_{li}}
		{\pi_i} I_i, \quad l=1,...,q;$$ and  $$\frac {\partial \widetilde {W}(\gamma)} {\partial \gamma_{y}}  = \sum_{i=1}^N {\frac{(\pi_i-(\gamma_v v_i+\gamma_y y_i)) y_i}{\pi_i}}I_i$$
where $I_i$ denotes the sampling indicator. Note that in the specified equations the observed values of the variables $Z$, $Y$ and $V$ are held fixed and the only source of randomness is expressed by the sampling indicators $I_1,...,I_N$, which can take the values $0$ and $1$.
\par 
Denote
		\begin{center}
		\begin{equation}
		\widetilde{J}=(\widetilde{J_1},...,\widetilde{J_q},{\widetilde{J}}_{q+1})^t=\left( \frac{\partial \widetilde{W}(\gamma)} {\partial \gamma_{v_1}}			,...,
		\frac {\partial \widetilde{W}(\gamma)} {\partial \gamma_{v_q}},
		\frac {\partial \widetilde{W}(\gamma)}{\partial \gamma_{y}}  \right)^t.
		\end{equation}
		\end{center}

		\begin{center}
		\begin{equation}
		J=(J_1,...,J_q,J_{q+1})^t=\left( \frac{\partial W(\gamma)} {\partial \gamma_{v_1}}			,...,
		\frac {\partial W(\gamma)}{\partial \gamma_{v_q}},
		\frac {\partial W(\gamma)}{\partial \gamma_{y}}  \right)^t
		\end{equation}
		\end{center}

Note that $$E_D(\widetilde{J_l} | \mathbf{\pi},\mathbf{y},\mathbf{v})  = 
\sum_{i=1}^N \frac {(\pi_i-(\gamma_v v_i+\gamma_y y_i)) v_{li}}{\pi_i} E_D(I_i)=  J_l= {0}_{(q+1) \times 1},$$
where $ {0}_{(q+1) \times 1}$ denotes a vector of zeros of dimension $q+1$, $E_D$ denotes the expectation over all possible sample selections, for the fixed population values, and $\mathbf{\pi},\mathbf{y},\mathbf{v}$ denote the vector of inclusion probabilities and the vectors of the population realizations of the variables $Y$and $V$correspondingly. Therefore, the following equations can be specified: 
		\begin{center}
		\begin{equation}
		\widetilde{J_l}=0_{(q+1) \times 1}+ \nu_l,\quad l=1,...,q+1
		\end{equation}
		\end{center}
where $\nu=(\nu_1 , \nu_2 ,...,\nu_{q+1})^t$ is a $q+1$-variate random variable. Implication of the equations (20) is that even were  $\gamma$ known, it is unlikely that the components of $\widetilde{J_l}$ would be equal to zero for any selected sample $S$ due to sampling variability, although we expect them to be close to zero. Noting that $\widetilde{J_l}$ are defined by sums of random variables, we assume that given the vector of unknown parameters $\gamma$, the vector of random variables $\nu$ can be approximated by a $q+1$-variate normal distribution, that is

		\begin{center}
		\begin{equation}
	 {\nu |\mathbf{\pi}_{obs},\mathbf{y}_{obs},\mathbf{v}_{obs}, \gamma} \approx {N (0_{(q+1) \times 1}, \Sigma(\gamma))}
		\end{equation}
		\end{center}
where $\mathbf{\pi}_{obs}$, $\mathbf{y}_{obs}$ and $\mathbf{v}_{obs}$ denote the observed parts of the vectors $\mathbf{z}$, $\mathbf{y}$ and $\mathbf{v}$ correspondingly.
A well known result of a classical sampling theory states that under the randomization distribution, the components of the matrix $\Sigma(\gamma)$ can be derived as follows:
		\begin{center}
		\begin{equation}
	  \left[\Sigma(\gamma)\right]_{k,l} = \sum_{i=1}^{N} \sum_{j=1}^{N} {\frac{\widetilde{e_i} \widetilde{e_j} v_{ ki }v_{ lj }}{ \pi_i \pi_j} (\pi_{ij}-\pi_i \pi_j)},
		\end{equation}
		\end{center}
where $\widetilde{e_i}=\pi_i-(\gamma_v v_i+\gamma_y y_i).$ \\
Then, based on the sample measurements, (22) can be estimated as
		\begin{center}
		\begin{equation}
	  \left[\hat{\Sigma}(\gamma)\right]_{k,l} = \sum_{i=1}^{n} \sum_{j=1}^{n} {\widetilde{e_i}
		\widetilde{e_j} v_{ ki }v_{ lj }} \left ( \frac{1}{\pi_i \pi_j}-\frac{1}{\pi_{ij}} \right)
		\end{equation}
		\end{center}
\begin{remark}[]
Calculation of the components of the variance matrix $\Sigma(\gamma)$, generally requires knowledge of the joint sampling probabilities for the units $k$ and $l$, where $k,l=1,...,N$. These probabilities are generally not specified, however, if they are proportional to a design variable, it can be shown that the estimator (23) can be rewritten as 

		%\begin{center}
		%\begin{equation}
	  $\left[\hat{\Sigma}(\gamma)\right]_{k,l} = \frac{(n-1)S_\pi}{n}\sum_{i=1}^{n} \sum_{j=1}^{n} \frac {\widetilde{e_i} \widetilde{e_j}
		 v_{ ki }v_{ lj }}{\pi_i+\pi_j} \left (\frac{\pi_i}{S_{\pi}-\pi_i}+\frac{\pi_j}{S_\pi-\pi_j} \right),$ 
		where $S_\pi=\frac{N}{n} \sum_{i=1}^n \pi_i$
		%\end{equation}
		%\end{center}
If sampling probabilities are proportional to some function of the design variable, calculation of $\Sigma(\gamma)$ will require knowledge of the design variable value for all sample units. 
\end{remark}
In order to apply a Bayesian approach we use the model (21) and a noninformative prior on  
$\gamma$, $h(\gamma)$. Then 

		\begin{center}
		\begin{equation}
	  f(\gamma | \mathbf{\pi}_{obs},\mathbf{y}_{obs},\mathbf{v}_{obs}) \propto \phi_{{0_{(q+1) \times 1} , \hat{\Sigma}(\gamma)}}(\widetilde{J} | 
		\mathbf{\pi}_{obs},\mathbf{y}_{obs},\mathbf{v}_{obs}; 		  \gamma) h(\gamma),
		\end{equation}
		\end{center}
where $ \phi_{{0_{(q+1) \times 1} , \hat{\Sigma}(\gamma)}}$ denotes a $q+1$-variate normal density function with the expectation vector, $0_{(q+1) \times 1}$ and the variance matrix equal to $\hat{\Sigma}(\gamma)$.   
%Now, let us consider inference based on the joint distribution over all possible sample outcomes, given the population values, and over all %possible realizations of the population outcomes. As mentioned before, a distributional assumption on the conditional distribution of $Z$ is %required. In what follows we assume that $\pi_i \sim N(v_i \gamma_v+\gamma_y y_i, \sigma_\pi^2)$, however the calculations can be modified to %the cases, discussed in Remark 1.  
%Let ${\nu | \mathbf{y}_{obs},\mathbf{v}_{obs}, \gamma} \approx N (M(\gamma), \Omega(\gamma))$. Then $$
\begin{remark}[]
Application of the proposed method to the situations where the sample selection mechanism follows model (6) is straightforward. In my research I would like to extend the proposed methodology to the cases, where the selection models have more complex forms.
\end{remark}
The regression parameters $\beta$, indexing the population model, can be estimated by implementation of the same approach, by defining the population sums of squares of the residuals as $U(\beta) = \sum_{i=1 }^N (y_i-x_i \beta)^2$ and their Horvitz-Tompson estimators,
$$\widetilde{U}(\beta) = \sum_{i=1 }^n\frac{ (y_i-  x_i \beta)^2}{\pi_i}$$.
\begin{remark}[]
The methods of estimation of unknown parameters $\beta$, hypothesis testing and identification of the model, holding in the population are being developed by the author at present. The author is also working on extending the proposed methodology to the situations, where the population model belongs to the family of GLM.
\end{remark}

\begin{remark}[]
An additional interesting extension of the proposed methodology is incorporation of constraints in the estimation process. This is also one of the direction of my future work. 
\end{remark}

\begin{remark}[]
As previously mentioned, we assume a flat prior on all unknown parameters. It is generally known that the posterior distributions may be heavily influenced by the priors. In this research I also intend to investigate and to implement various methods of sensitivity analysis. 
\end{remark}

\subsection{Application of the FBST}
In this section we consider a simple case of hypothesis testing under the model defined by (14), $H: \gamma_y=0$, where rejection of $H$ implies that the sample selection mechanism is ignorable. Extension to testing hypothesis of the form $\widetilde{H}: \widetilde\gamma_y=0$, where  
$Span(\gamma  \setminus  \widetilde{\gamma}_y) \subseteq  Span(\gamma)$ for more complex sample selection models is straightforward. Denote by ${\hat{\gamma}}^{post}$ the random draws from the posterior distribution $f( \gamma | \mathbf{\pi}_{obs},\mathbf{y}_{obs},\mathbf{v}_{obs})$, obtained by application of the MCMC to the full model, and let $\dim({\hat{\gamma}}^{post})=K \times (q+1)$. Denote by $\gamma_0$ the vector of unknown parameters, indexing the sampling selection model under $H$.
Let
		\begin{center}
		\begin{equation}
		{\hat{\gamma}}_0=\underset{H} {\mathrm{argmax}}~
		\phi_{0_{(q+1) \times 1)} , \hat{\Sigma}(\gamma)}
		(\widetilde{J} | \mathbf{\pi}_{obs},\mathbf{y}_{obs},\mathbf{v}_{obs}; \gamma) h(\gamma)
		\end{equation}
		\end{center}
and 
		\begin{center}
		\begin{equation}
		T= \left\{ \gamma: f(\gamma | \mathbf{\pi}_{obs},\mathbf{y}_{obs},\mathbf{v}_{obs}) >f( {\hat{\gamma}}_0 | 
		\mathbf{\pi}_{obs},\mathbf{y}_{obs},\mathbf{v}_{obs}) \right\}
		\end{equation}
		\end{center}
It follows (see Reference) then that derivation of the value $\overline{ev}$, requires computation of the probability
$P(T |  \mathbf{\pi}_{obs},\mathbf{y}_{obs},\mathbf{v}_{obs})$. This probability can be estimated, using the posterior draws, as follows
${\hat{\gamma}}^{post}$. 
		\begin{center}
		\begin{equation}
		P( T | \mathbf{\pi}_{obs},\mathbf{y}_{obs},\mathbf{v}_{obs}) \approx \frac{1}{K} \sum_{k=1}^K I_{\left\{ \frac{f(\gamma_k |
		\mathbf{\pi}_{obs},\mathbf{y}_{obs},\mathbf{v}_{obs})} {f({\hat{\gamma}}_0 | \mathbf{\pi}_{obs},\mathbf{y}_{obs},\mathbf{v}_{obs})}>1 \right\}},
		\end{equation}
		\end{center}
where
		\begin{center}
		\begin{equation}
		\frac {f(\gamma_k | \mathbf{\pi}_{obs},\mathbf{y}_{obs},\mathbf{v}_{obs})} 
		{f({\hat{\gamma}}_0 | \mathbf{\pi}_{obs},\mathbf{y}_{obs},\mathbf{v}_{obs})}=
		\frac{\phi_{0_{((q+1) \times 1)}, \hat{\Sigma}(\gamma_k)}(\widetilde{J} | \mathbf{\pi}_{obs},\mathbf{y}_{obs},\mathbf{v}_{obs}; \gamma_k) 
		h(\gamma_k)}
		{\phi_{0_{(q \times 1)}, \hat{\Sigma}({\hat{\gamma}}_0)}(\widetilde{J} | \mathbf{\pi}_{obs},\mathbf{y}_{obs},\mathbf{v}_{obs}; 
		{\hat{\gamma}}_0) h({\hat{\gamma}}_0)}
		\rho(\mathbf{\pi}_{obs},\mathbf{y}_{obs},\mathbf{v}_{obs})
		\end{equation}
		\end{center}
and 
		\begin{center}
		\begin{equation}
		\rho(\mathbf{\pi}_{obs},\mathbf{y}_{obs},\mathbf{v}_{obs})=
		\frac{\int \phi_{\mathbf{0}_{((q+1) \times 1)}, \hat{\Sigma}(\gamma)}(\widetilde{J} | \mathbf{\pi}_{obs},\mathbf{y}_{obs},\mathbf{v}_{obs}; \gamma) 
		h(\gamma ) d\gamma} 
		{\int {\phi_{\mathbf{0}_{(q \times 1)}, \hat{\Sigma}(\gamma_0)}(\widetilde{J} | \mathbf{\pi}_{obs},\mathbf{y}_{obs},\mathbf{v}_{obs}; \gamma_0) 
		h(\gamma_0)d\gamma_0}}
		\end{equation}
		\end{center}
Therefore, application of the FBST requires two following steps:
\begin{enumerate}
\item A maximization step (25).
\item Computation of the ratio of the integrals, defined in (29). 
\end{enumerate}
The first step constitutes a standard maximization problem, which can usually be carried out by application of a standard multivariate Newton-Raphson algorithm, and therefore does not require intensive computations. The second step involves more complex computations, however, it can be carried out by application of an importance sampling, using Monte Carlo techniques, as described by Zacks and Stern (2003). The authors also derive the sample size, which is required for attaining the desired precision of the integrals ratio.  
\begin{remark}[]
One of the objectives of my research at this point is to approximate the value of $\rho(\mathbf{\pi}_{obs},\mathbf{y}_{obs},\mathbf{v}_{obs})$ in order to avoid computation of the integrals in (29). This will significantly simplify the process of calculation of the evidence value. 
\end{remark}

\section{Simulation Study}
In order to test the proposed approach and to compare its performance to some other approaches, described in a literature review section, we performed a small simulation study, which consisted of a few experiments. For each experiment we generated $M=200$ populations of size $N=500$, and for each generated population we selected one sample of size $n=50$, such that the units were randomly selected with the inclusion probabilities proportional to the values of the design variable $Z$. We used the following population and sample selection models. 
\begin{equation}
Y_i=3.5+0.8x_i-0.1v_i+\epsilon_i, i=1,...,N
\end{equation}
where $\epsilon_i \sim N(0,1.5)$ and the auxiliary variables $X$ and $V$ were generated from
$Gamma(1,1)$ and $Poisson(3)$ distributions correspondingly. The true model for the design variable, $Z$ is: 
\begin{equation}
Z_i=4+2.5v_i+0.15y^2_i+\nu_i, 
\end{equation}
where $\nu_i \sim N(0,2.5)$. 
The primary objective of the simulation study is to identify the model, holding for $E(Z_i | v_i,y_i)$ by testing the following hypotheses.   
\begin{changemargin}{-0.5cm}{-0.5cm} 
\begin{enumerate}
\item $H: E(Z_i | v_i,y_i)=\gamma_0^{(0)}+\gamma_1^{(0)}v_i  \quad$  \textbf{vs}
 $\quad E(Z_i | v_i,y_i)=\gamma_0^{(1)}+\gamma_1^{(1)}v_i+\gamma_2^{(1)}y_i$
\item $H: E(Z_i | v_i,y_i)=\gamma_0^{(1)}+\gamma_1^{(1)}v_i+\gamma_2^{(1)}y_i \quad$ \textbf{vs} $\quad E(Z_i | v_i,y_i)=\gamma_0^{(2)}+\gamma_1^{(2)}x_i+\gamma_2^{(2)}y_i+\gamma_3^{(2)}y^2_i$
\item $H: E(Z_i | v_i,y_i)=\gamma_0^{(1)}+\gamma_1^{(1)}v_i+\gamma_2^{(1)}y_i  \quad$  \textbf{vs}
 $\quad E(Z_i | v_i,y_i)=\gamma_0^{(3)}+\gamma_1^{(3)}v_i+\gamma_2^{(3)}y_i^2$
\end{enumerate}
\end{changemargin}
For each experiment, $j=1,2,3$ and each sample $i, i=1,...,M$ we compute the value of the Bayesian evidence value in favour of H, ${ev}_{ji}$ as presented in Section 4.2, using a normal flat prior. As mentioned above the FBST rejects $H$ whenever $ev$ is small. In order to define a rejection region, we utilize the asymptotic distribution of the evidence value under $H$, provided by Pereira et al. (2008). The tables below summarize proportions of the samples, where the hypothesis $H$ was rejected for various significance level. We illustrate the results for the proposed test and the classical Likelihood Ratio test (LR), based on the model (21). For the first experiment we also applied the test proposed by Pfeffermann and Sverchkov (1999), which is based on the identity in (11). As mentioned above this test can only be implemented to testing informativeness of the sampling selection mechanism and is not applicable if both tested models are informative. We denote by $PS(1)$ and by $PS(2)$ the tests based on the correlations between the sampling weights and the first and the second powers of the residuals, as described in the end of the Section 3.1.  

\begin{table}[h]
\centering
\caption{\textbf{Proportions of samples, where H was rejected.}}
\label{1}
\begin{tabular}{ |c| c c c c| c c| c c |}
\hline
\multicolumn{1}{|c|}{} & \multicolumn{4}{|c|}{Experiment 1} & \multicolumn{2}{|c|}{Experiment 2} & \multicolumn{2}{|c|}{Experiment 3}\\

Significance level & FBST & LR & PS(1) & PS(2) & FBST & LR & FBST & LR\\ 
\hline
 0.010 & 0.921 & 0.908 & 0.910 & 0.011 & 0 & 0.116 & 0.955 & 0.928 \\  
%\hline 
0.025 & 0.966 & 0.940 & 0.955 & 0.017 & 0.007 & 0.176 & 0.980 & 0.952 \\ 
%\hline 
0.050 & 0.977 & 0.960 & 0.977 & 0.045 & 0.062 & 0.284 & 0.980 & 0.964 \\  
%\hline 
0.100 & 0.989 & 0.984 & 0.977 & 0.119 & 0.205 & 0.396 & 0.995 & 0.986 \\
\hline  
\end{tabular}
\end{table}

The table above illustrates that the FBST prsents very good power properties for the first and the third experiments. The high probability of rejection of $H$ in the first experiment implies that our method succeeds in revealing that the sampling mechanism is indeed informative (recall that the true model for the sampling selection process included the quadratic term of the value of outcome variable). In the third experiment both specified models are informative. In this case the proposed method shows good performance in revealing the correct model. For the second experiment, both models are not correct. The results indicate in this case, that both methods generally do not reject the model with a smaller number of coefficients.      

In order to validate the proposed methods, we carried out an additional experiment. For this experiment the tested models are the same as in the first experiment, however the values of the design variable $Z$ were generated under the model under the hypothesis $H$ of this experiment. We expect that the proportion of the samples where the hypothesis $H$ is rejected will be close to the significance level.   
\begin{table}[h]
\centering
\caption{\textbf{Empirical and theoretical distribution of test statistics under model $H$.}}
\label{1}
\begin{tabular}{ |c| c c c c|}
\hline
%\multicolumn{1}{|c|}{} & \multicolumn{4}{|c|}{Experiment 1} \\

Nominal level & FBST & LR & PS(1) & PS(2) \\ 
\hline
0.025 & 0.014 & 0.032 & 0.033 & 0.033\\ 
0.050 & 0.037 & 0.056 & 0.044 & 0.061 \\  
0.100 & 0.074 & 0.096 & 0.083 & 0.122 \\
0.250 & 0.200 & 0.228 & 0.200 & 0.194 \\
0.500 & 0.510 & 0.464 & 0.510 & 0.467 \\
0.750 & 0.749 & 0.732 & 0.778 & 0.778 \\
0.900 & 0.900 & 0.896 & 0.883 & 0.894 \\
0.950 & 0.946 & 0.936 & 0.939 & 0.939 \\

\hline  
\end{tabular}
\end{table}

In general, the empirical distribution of all the statistics is sufficiently close to the nominal values, thus validating the use of all the discussed methods.

\providecommand{\href}[2]{#2}\begingroup

\end{document}